\documentclass[aps,pre,twocolumn,floats]{revtex4}
\usepackage{epsfig}
\begin{document}

\newcommand{\tbox}[1]{\mbox{\tiny #1}}
\newcommand{\half}{\mbox{\small $\frac{1}{2}$}}
\newcommand{\mbf}[1]{{\mathbf #1}}

%%%%%%%%%%%%%%%%%%%%%%%%%%%%%%%%%%%%%%%%%%%%%%%%%%%%%%%%%%%%%%%

\title{Quantum irreversibility, perturbation independent decay,
and the parametric theory of the local density of states}

\author{Diego A. Wisniacki$^1$ and Doron Cohen$^2$}

\date{October 2001, May 2002,August 2002}

\affiliation{
$^1$\mbox{Departamento de F\'{\i}sica,
Comisi\'on Nacional de Energ\'{\i}a At\'omica.
Av. del Libertador 8250,
1429 Buenos Aires, Argentina.} \\
$^2$\mbox{Department of Physics,
Ben-Gurion University,
Beer-Sheva 84105, Israel}
}

%%%%%%%%%%%%%%%%%%%%%%%%%%%%%%%%%%%%%%%%%%%%%%%%%%%%%%%%%%%%%%%

\begin{abstract}
The idea of perturbation independent decay (PID)
has appeared in the context of survival-probability
studies, and lately has emerged in the context
of quantum irreversibility studies.  In both cases
the PID reflects the Lyapunov instability of the underlying
semiclassical dynamics, and it can be distinguished
from the Wigner-type decay that holds in the
perturbative regime. The theory of the survival
probability is manifestly related to the parametric
theory of the local density of states (LDOS).
In contrast to that the physics of quantum irreversibility
requires to take into account subtle cross correlations
which are not captured by the LDOS alone.
\end{abstract}

\maketitle

%%%%%%%%%%%%%%%%%%%%%%%%%%%%%%%%%%%%%%%%%%%%%%%%%%%%%%%%%%%%%%
\section{Introduction}

The study of quantum irreversibility \cite{peres} has become lately
of much interest \cite{jalabert,philip,casati,tomsovic,prosen,wisniacki,rvs}
due to its potential relevance to quantum computing,
and to the theory of dephasing \cite{wld,dsp}.
Following \cite{peres} we define in Section~2 the main
object of the present Paper: This is the "fidelity",
also known as "Loschmidt echo",
which constitutes a measure for quantum irreversibility.

The analysis of "fidelity" necessitates a generalization
of the theory regarding "survival probability" \cite{heller}.
In Section~3 we remind the reader that the latter reduces
to the analysis of the local density of states (LDOS) \cite{wls}.
Is it possible to make a similar reduction in case of the "fidelity"?
At first sight such reduction looks feasible because the
general physical picture looks very similar (Sections~5-6).

In the present Paper we claim (Section~6),
and prove by a numerical example (Sections~7-8),
that the study of fidelity cannot be reduced
to analysis of LDOS functions.
Rather, it is essential to take into account
subtle cross correlations which are not captured
by the LDOS alone.

The object of the present study is common to almost
all the "quantum chaos" studies. Namely, to figure out
what is the role of semiclassical mechanics
in quantum mechanics.  Whenever we find such
(semiclassical) "fingerprints",
we call them "non-universal effects".
Most of the studies in the "quantum chaos"
literature during the last 20 years have been
devoted to figuring out the non-universal features
of the energy spectrum. The main tool in singling
out such features is a comparison with
the predictions of random matrix theory (RMT).

In the present Paper we use the same "philosophy".
Namely, we identify "non universal effects"
by making a comparison with a corresponding
random matrix model. On the other hand
we discuss (Sections 10-11) a unifying theoretical
picture that put the study of "quantum irreversibility"
in the larger context of phase-space based
semiclassical approach.

An important ingredient in the understanding of
"non universal features" follows from studies
of the the clash between "perturbation theory",
semiclassical theory, and RMT \cite{dsp,wls,crs}.
A major realization is that semiclassical theory
and RMT lead to {\em different} non-perturbative limits.
Hence the resolution of the clash between the
different theories involves the identification
of different {\em regimes} of behavior.
This applies in general to the analysis
of time dependent dynamics \cite{crs},
and in particular to the analysis of wavepacket dynamics,
decay of the survival probability,
structure of the LDOS \cite{wls}, and naturally
also to "quantum irreversibility" studies.

Specifically, we distinguish in the
present Paper between regimes of "perturbative"
and "non-universal" behavior,
and we define and study a billiard
related model, where we have full control over
the "borders" between these regimes.
The conclusions are summarized in Section~12.

%%%%%%%%%%%%%%%%%%%%%%%%%%%%%%%%%%%%%%%%%%%%%%%%%%%%%%%%%%%%%%
\section{The "fidelity"}

Consider a system whose evolution is governed
by the chaotic Hamiltonian
\begin{eqnarray} \label{eq1}
{\cal H}={\cal H}(Q,P;x)
\end{eqnarray}
where $(Q,P)$ is a set of canonical coordinates,
and $x$ is a parameter. Later (Section~7) we are going
to consider, as an example, a billiard system,
where $(Q,P)$ are the position and the momentum
of a particle, while $x$ is used in order to parametrize
the shape of the billiard. Specifically, for a stadium
we define $x$ as the length of the straight edge,
and adjust the radius parameter such that the total
area is kept constant.

Consider some ${\cal H}_0={\cal H}(Q,P;x_0)$,
and define $\delta x = x-x_0$. Assume that $\delta x$
is classically small, so that both ${\cal H}_0$ and ${\cal H}$
generate classically chaotic dynamics of similar nature.
Physically, going from ${\cal H}_0$ to ${\cal H}$ may
signify a small change of an external field.
In case of billiard system $\delta x$ parametrizes the
displacement of the walls.
Given a preparation $\Psi_0$,
the fidelity is defined as \cite{rmrk0}:
\begin{eqnarray} \label{e1}
M(t;\delta x) \ &=& \ |m(t;\delta x)|^2 \\
m(t;\delta x) \  &\equiv& \  \langle \Psi_0 |
\exp(+i{\cal H}t) \exp(-i{\cal H}_0 t)
| \Psi_0 \rangle
\end{eqnarray}
If $\Psi_0$ is an eigenstate $|E_0\rangle$ of ${\cal H}_0$,
then $M(t;\delta x)$ is equal to the survival
probability $P(t;\delta x)$, which is defined as
\begin{eqnarray}
P(t;\delta x) \ &=& \ |c(t;\delta x)|^2 \\
c(t;\delta x) \ &\equiv& \langle E_0 |
\exp(-i{\cal H}t) | E_0 \rangle
\end{eqnarray}
In the general case the preparation $\Psi_0$
does {\em not} have to be an eigenstate of ${\cal H}_0$.
To be specific one assumes that $\Psi_0$ is
a Gaussian wavepacket. It is now possible
to define a different type of survival probability
as follows:
\begin{eqnarray}
P(t;\mbox{wpk}) \ &=& \ |c(t;\mbox{wpk})|^2 \\
c(t;\mbox{wpk}) \ &\equiv& \langle \Psi_0 |
\exp(-i{\cal H}t) | \Psi_0 \rangle
\end{eqnarray}
We assume that $\delta x$ is small enough
so that we do not have to distinguish
between ${\cal H}$ and ${\cal H}_0$ in the
latter definition.

One may regard $\Psi_0$ as an eigenstate
of some preparation Hamiltonian ${\cal H}_{\tbox{wpk}}$.
Specifically, if $\Psi_0$ is a Gaussian wavepacket,
then it is the groundstate of Hamiltonian
of the type $(P-P_0)^2+(Q-Q_0)^2$
that differs enormously from~${\cal H}$.
Thus we have in the general case three Hamiltonians:
\begin{itemize}
\setlength{\itemsep}{0cm}
\item The preparation Hamiltonian ${\cal H}_{\tbox{wpk}}$
\item The unperturbed evolution Hamiltonian  ${\cal H}_0$
\item The perturbed evolution Hamiltonian ${\cal H}$
\end{itemize}
Above we have distinguished between two cases:
The relatively simple case where
${\cal H}_{\tbox{wpk}}={\cal H}_0$.
And the more general case,
where we assume that the difference
$||{\cal H}_{\tbox{wpk}}-{\cal H}_0||$
is in fact much larger compared with
the perturbation $||{\cal H}-{\cal H}_0||$.
The strength of the perturbation is controlled
by the parameter $\delta x$.

%%%%%%%%%%%%%%%%%%%%%%%%%%%%%%%%%%%%%%%%%%%%%%%%%%%%%%%%%%%%%%
\section{The LDOS functions}

Consider first the special case
where ${\cal H}_{\tbox{wpk}}={\cal H}_0$.
In such case the fidelity amplitude $m(t;\delta x)$
is just the Fourier transform
of the local density of states (LDOS):
\begin{eqnarray} \label{e3}
\rho(\omega;\delta x) = \sum_n
|\langle n(x)|E_0 \rangle|^2
\ \delta(\omega-(E_n(x)-E_0))
\end{eqnarray}
For technical reasons, we would like
to assume that there is an implicit
average over the reference state $|E_0\rangle$.
This will enable a meaningful comparison
with the more general case which is discussed below.

In the general case,
where ${\cal H}_{\tbox{wpk}} \ne  {\cal H}_0$,
one should recognize the need in defining
an additional LDOS function:
\begin{eqnarray} \label{e4}
\rho(\omega;\mbox{wpk}) = \sum_n
|\langle n|\Psi_0 \rangle|^2
\ \delta(\omega-(E_n-E_0))
\end{eqnarray}
In this context $E_0$ is consistently re-defined
as the mean energy of the wavepacket.
Recall again that $\delta x$ is assumed
to be small enough, so that we do not have
to distinguish between ${\cal H}$ and ${\cal H}_0$
in the latter definition.

The Fourier transform of $\rho(\omega;\mbox{wpk})$
equals (up to a phase factor) to the survival
amplitude $c(t;\mbox{wpk})$ of the wavepacket.
The physics of $c(t;\mbox{wpk})$ is
assumed to be of "semiclassical" type.
We shall define what do we mean by
"semiclassical" later on. The same notion
is going to be used regarding $c(t;\delta x)$
if the perturbation ($\delta x$) is large enough.

%%%%%%%%%%%%%%%%%%%%%%%%%%%%%%%%%%%%%%%%%%%%%%%%%%%%%%%%%%%%%%
\section{Definitions of $\Gamma$ and $\gamma$}

In this Paper we measure the "width"
of the LDOS of Eq.(3) in energy units \cite{rmrk0},
and denote it by $\Gamma(\delta x)$. 
A practical numerical definition of $\Gamma(\delta x)$ 
is as the width of the central region that 
contains $70\%$ of the probability.
This corresponds to the notion of "core width" in \cite{wls}.
If $\delta x$ is not too large (see definition
of "Wigner regime" in the next section), one observes that
\begin{eqnarray}
\Gamma(\delta x) \ \propto \ \delta x^{2/(1{+}g)}
\end{eqnarray}
The value $g\sim 0$ applies for strong chaos \cite{wls,prm},
and it is the same as in Wigner's random matrix theory (RMT)
\cite{wigner}. In general (eg see Appendix) we can have $0<g<1$.
In fact, the value $g\sim 1$ applies to our numerical
model, which will be defined in Section~7.

The decay rate of either the fidelity or of the survival
probability (depending on the context) is denoted
by $\gamma(\delta x)$. The semiclassical value of the decay
rate, which is determined via a "wavepacket dynamics"
phase-space picture \cite{heller},
is denoted by $\gamma_{\tbox{scl}}$.
The Lyapunov exponent is denoted by $\gamma_{\tbox{cl}}$.

In order to determine $\gamma(\delta x)$
numerically one should plot $M(t;\delta x)$
against $t$, for a range of $\delta x$ values.
In Section~7 we are going to
define some model Hamiltonians for which
we have done simulations. These are called
\begin{itemize}
\setlength{\itemsep}{0cm}
\item LBH: Linearized Billiard Hamiltonian
\item RLBH: Randomized version of LBH.
\item MBH: Modified Billiard Hamiltonian
\item RMBH: Randomized version of MBH.
\end{itemize}
Fig.1a displays the results of the MBH simulations.
We see that the MBH decay is well approximated
by exponential function (Fig.~1a).
The dependence of the decay rate $\gamma_{\tbox{MBH}}$
on $\delta x$ is presented in Fig.2.
The RMBH decay (Fig.~1b) is badly approximated
by exponential function, but in order to make
a comparison we still fit it to exponential.
This is done in order to have quantitative
measure for the decay time.
Thus we have also $\gamma_{\tbox{RMBH}}(\delta x)$.

In Fig.2 we also plot the LDOS width $\Gamma(\delta x)$
as a function of $\delta x$ for the two models.
As far as $\Gamma(\delta x)$ is concerned the
two models are practically indistinguishable.
The inset contains plots of $\Gamma(\delta x)$
for the other two models (LBH,RLBH).
In later sections we shall discuss the significance
of the observed numerical results.

%%%%%%%%%%%%%%%%%%%%%%%%%%%%%%%%%%%%%%%%%%%%%%%%%%%%%%%%%%%%%%
\section{The decay of $P(t;\delta x)$}

The theory of the survival amplitude is on firm
grounds thanks to the fact that it is the Fourier transform
of the LDOS. According to \cite{wls}
there are three generic $\delta x$ regimes of behavior:
\begin{itemize}
\setlength{\itemsep}{0cm}
\item The standard perturbative regime.
\item The Wigner (or Fermi Golden Rule) regime.
\item The non-universal (semiclassical) regime.
\end{itemize}

In the standard perturbative regime
($\delta x \ll \delta x_c$)
the LDOS function Eq.(\ref{e3})
is predominantly a Kronecker delta.
This characterization constitutes
a {\em definition} of this regime.
For estimate of $\delta x_c$ in case
of billiards see Appendix.
The survival amplitude is obtained
via a Fourier transform of the Kronecker delta
dominated LDOS function. This leads to
a non-averaged $m(t;\delta x)$ that does not decay.
On the other hand the $E_0$ averaged $m(t;\delta x)$
has a Gaussian decay. The latter follows from the
observation \cite{peres} that the first order
correction $E_0(x)-E_0(x_0)$ has
typically a Gaussian distribution.

For intermediate values of $\delta x$ the decay
of $P(t;\delta x)$ is typically of exponential type with
\begin{eqnarray} \label{e_Wg}
\gamma=\Gamma(\delta x)/\hbar
\end{eqnarray}
This is known as Wigner-type (or as Fermi golden rule) decay.
It is a reflection of the Lorentzian-like line shape
of the LDOS function.
However, for large $\delta x$ we get into a non-universal
(semiclassical) regime, where we can apply the "wavepacket dynamics"
picture of \cite{heller}. Thus we find a semiclassical decay with
\begin{eqnarray}
\gamma = \gamma_{\tbox{scl}}
\end{eqnarray}
The Wigner regime, where Eq.(\ref{e_Wg}) holds,
is determined \cite{wls} by the condition
\begin{eqnarray} \label{eq_cond}
\Gamma(\delta x) \ll \hbar \gamma_{\tbox{scl}}
\end{eqnarray}
This inequality can be re-written as
$\delta x < \delta x_{\tbox{NU}}$.
The elimination defines a non-universal
(system specific) parametric scale
$\delta x_{\tbox{NU}}$.

In the non-universal regime the width of the LDOS
is semiclassically determined \cite{wls}.
In typical cases the width of the LDOS is proportional
to the strength of the perturbation, hence
\begin{eqnarray}
\gamma_{\tbox{scl}} \propto \delta x
\end{eqnarray}
But in some exceptional cases $\gamma_{\tbox{scl}}$
becomes perturbation independent.
Specifically, for billiard systems $1/\gamma_{\tbox{scl}}$
roughly equals to the mean time between collisions,
so we can write
\begin{eqnarray}
\gamma_{\tbox{scl}} \approx \gamma_{\tbox{cl}}
\end{eqnarray}
It is important to realize that the perturbation
independent decay (PID) of $c(t;\delta x)$ in
billiard systems, is a reflection of the $\delta x$
independence of LDOS function $\rho(\omega;\delta x)$
in the non-universal regime.
See \cite{prm} for a numerical study.

%%%%%%%%%%%%%%%%%%%%%%%%%%%%%%%%%%%%%%%%%%%%%%%%%%%%%%%%%%%%%%
\section{The decay of $M(t;\delta x)$}

A mature theory of fidelity is still lacking.
However, it has been realized in \cite{philip,casati}
that the same physical picture as in \cite{wls} arises:
For very small $\delta x$ we have Gaussian decay
[which corresponds to the $E_0$ averaged decay of the
survival amplitude]. For intermediate values of $\delta x$
we have Wigner-type decay with $\gamma=\Gamma(\delta x)/\hbar$.
For large $\delta x$ we enter into the semiclassical regime
where one finds "Lyapunov decay" \cite{bruno} with
$\gamma \approx   \gamma_{\tbox{scl}} \approx  \gamma_{\tbox{cl}}$.

In complete analogy with the case of survival
probability studies we can define (via Eq.(\ref{eq_cond}))
an analogous parametric scale \cite{philip} that
will be denoted by $\delta x_{\tbox{NUD}}$.
The semiclassical value ($\gamma_{\tbox{scl}}$)
of $\gamma$ is not necessarily the same for
$P(t;\delta x)$ and for $M(t;\delta x)$.
Therefore in general $\delta x_{\tbox{NU}}$
and $\delta x_{\tbox{NUD}}$ are not necessarily identical.

For simple shaped billiard system the $\gamma_{scl}$
of the survival probability, the $\gamma_{scl}$
of the fidelity, and the Lyapunov exponent $\gamma_{cl}$
are all equal to the inverse of the mean collision time.
The perturbation parameter $\delta x$ is defined
as the displacement of the billiard wall.
In the Appendix we derive the following result:
\begin{eqnarray}  \label{eq_DeB}
\delta x_{\tbox{NUD}} \sim  \delta x_{\tbox{NU}} \sim 2\pi/k
\end{eqnarray}
where $2\pi/k$ is the De-Broglie wavelength
of a particle with mass $m$, corresponding
to the kinetic energy $E_0 = (\hbar k)^2/2m$.
This results holds for hard walled billiard.

In the following we want to demonstrate
the distinction between $\delta x_{\tbox{NUD}}$
and $\delta x_{\tbox{NU}}$. Therefore we consider
a modified billiard Hamiltonian (MBH) for which
\begin{eqnarray}
\delta x_{\tbox{NUD}} \ll \delta x_{\tbox{NU}}
\end{eqnarray}
The above inequality reflects the general case,
in which the $\gamma_{scl}$ of $M(t;\delta x)$ is
different (smaller) from the $\gamma_{scl}$ of $P(t;\delta x)$.

The fact that $P(t;\delta x)$ is a special case
of $M(t;\delta x)$, and the fact that similar ideas
(semiclassical decay versus Wigner-type decay)
have emerged in the latter case, naturally suggests that
the same physics is concerned.
If it were really the "same physics", it would imply that
the main features of $M(t;\delta x)$ are determined
by a simple minded theory that involves
the LDOS function $\rho(\omega;\delta x)$
in some combination
with the LDOS function $\rho(\omega;\mbox{wpk})$.
It is the purpose of the following sections
to demonstrate that a simple-minded theory is not enough.
The semiclassical PID in the case of $M(t;\delta x)$
necessitates a non-trivial extension
of the LDOS parametric theory.

%%%%%%%%%%%%%%%%%%%%%%%%%%%%%%%%%%%%%%%%%%%%%%
\section{Definition of the model}

Our model Hamiltonian is
the linearized billiard Hamiltonian (LBH)
of a stadium system \cite{wisniacki}.
It can be written as
\begin{eqnarray}
{\cal H} = \mbf{E}+\delta x\mbf{B}
\end{eqnarray}
Here $\mbf{E}$ is the ordered diagonal matrix $\{ E_n(x_0) \}$.
The eigen-energies of the quarter stadium billiard, with
straight edge $x_0=1$, have been determined numerically.
The perturbation due to $\delta x$ deformation,
is represented by the matrix $\mbf{B}$.
Also this matrix has been determined
numerically as explained in \cite{wisniacki}.

In the following numerical study we have 
considered not the LBH, but rather 
a modified Billiard Hamiltonian (MBH), 
which is obtained from the LBH by the replacement 
\begin{eqnarray}
\mbf{B}_{nm} \ \mapsto \ G(n-m)\times\mbf{B}_{nm}
\end{eqnarray}
where $G(n-m)$ is a Gaussian cutoff function. 
This corresponds physically to having soft 
walls (for explanation of this point 
see Appendix~J of \cite{frc}).
It is important to realize that the "exact" 
physical interpretation of either the LBH 
(as an approximation for the Billiard Hamiltonian), 
or the MBH (as a soft wall version of LBH), 
is of no importance for the following.
The LBH and the MBH are both mathematically 
"legitimate"  Hamiltonians.

In the next section we explain the numerical strategy
which we use in order to prove our main point.
This incorporates the random matrix theory (RMT)
strategy which has been applied in \cite{wpk}
in order to demonstrate that the semiclassical theory
and RMT lead to {\em different non-perturbative limits}.
The randomized LBH (RLBH) is obtained
by sign-randomization of the off-diagonal
elements of the $\mbf{B}$ matrix:
\begin{eqnarray}
\mbf{B}_{nm} \ \mapsto \ \pm \mbf{B}_{nm}
\ \ \ \ \mbox{(random sign)}
\end{eqnarray}
The randomized MBH (RMBH) is similarly defined.
The purpose in making comparison with "randomized" 
Hamiltonian, is the ability to distinguish between  
"universal" and "non-universal" effects. 
Making such distinction is a central theme 
in the "quantum chaos" literature. Usually 
such "comparisons" are made in the context of 
spectral statistics analysis, while here,
following \cite{wpk} we are doing this "comparison"
in the context of quantum dynamics analysis.

%%%%%%%%%%%%%%%%%%%%%%%%%%%%%%%%%%%%%%%%%%%%%%
\section{The numerical study}

The first step of the numerics is to
calculate the "width" $\Gamma(\delta x)$,
of the LDOS function $\rho(\omega;\delta x)$.
We know from previous studies \cite{wls,prm} that
for hard-walled billiard system $\Gamma(\delta x)$
shows semiclassical saturation
for $\delta x > \delta x_{\tbox{NU}}$, where
$\delta x_{\tbox{NU}}$ roughly equals to De-Broglie
wavelength (Eq.(\ref{eq_DeB})).
This implies PID for the survival probability.
With the LBH  we still see (inset of Fig.2)
a reminisces of this saturation.
Note that $k\sim50$ and hence $\delta x_{\tbox{NU}}\sim 0.1$.
In contrast to that, with the RLBH there is
no indication for saturation.
This implies that non-trivial correlations
of off-diagonal elements play an essential role
in the parametric evolution of the LDOS.
(See \cite{rmrkP} regarding terminology).

By {\em modifying} the billiard Hamiltonian we are able
to construct an artificial model Hamiltonian (MBH) where
the two parametric scales are well separated
($\delta x_{\tbox{NUD}} \ll \delta x_{\tbox{NU}}$).
Thus within a large intermediate $\delta x$ range \cite{rmrk1}
we do not have PID for $P(t;\delta x)$,
but we still find PID for $M(t;\delta x)$. See Fig.2.

In order to prove that the observed PID is not
a trivial reflection of $\rho(\omega;\mbox{wpk})$
we have defined the associated "randomized" Hamiltonian (RMBH).
The LDOS functions (\ref{e3}) and (\ref{e4}) are practically
{\em not} affected by the sign-randomization procedure: 
the sign-randomization procedure has almost
no effect on $\Gamma(\delta x)$.
In spite of this fact we find that
the previously observed PID of $M(t;\delta x)$ goes away:
we see (Fig.2) that for the MBH there is no
longer PID in the relevant $\delta x$ range \cite{rmrk1}.
This indicates that the PID was of semiclassical "off-diagonal" origin.

We see that both qualitatively and quantitatively the
sign-randomization procedure has a big effect on $M(t;\delta x)$.
Therefore, we must conclude that the correlations of the
off-diagonal terms is still important for the physics of
$M(t;\delta x)$. This holds in spite of the fact
that the {\em same} off-diagonal correlations are not
important for the LDOS structure. This implies that
the theory of $M(t;\delta x)$ necessitates a non-trivial
extension of the parametric LDOS theory.

%%%%%%%%%%%%%%%%%%%%%%%%%%%%%%%%%%%%%%%%%%
%%%%%%%%%%%%%%%%%%%%%%%%%%%%%%%%%%%%%%%%%%
\section{The simple minded theory}

The purpose of the present section is
to explain what type of "fidelity physics" can
be obtained if we do not take non-universal
(semiclassical) features of the dynamics into
account. Such theory is expected to be valid
in case of RMT models.
Let $\rho_{\tbox{eff}}(\omega;\delta x)$
be the Fourier transform of $m(t;\delta x)$.
It can be written as
\begin{eqnarray} \label{e5}
\rho_{\tbox{eff}}(\omega;\delta x) =
\sum_{\omega'} f(\omega') \delta(\omega-\omega')
\end{eqnarray}
where the summation is over energy differences
$\omega'=(E_n(x)-E_m(x_0))$, and $f(\omega')$
is a product of the overlaps $\langle n(x)|m(x_0)\rangle$,
and $\langle m(x_0) | \Psi_0 \rangle$
and $\langle \Psi_0 | n(x) \rangle$.
It is clear that $f(\omega')$ satisfies the sum rule
$\sum_{\omega} f(\omega) =  1$.
On the other hand, if the number of principle
components (participation ratio) of the LDOS is $N$,
then the sum over $|f(\omega)|$ gives $N^{1/2}$.
Thus we conclude that $f(\omega)$ should have
random-like phase (or random-like sign) character.
Therefore, if we ignore the system specific features,
we can regard $f(\omega)$ as the Fourier components
of a noisy signal. These Fourier components satisfy
\begin{eqnarray} \label{e7}
\langle f(\omega) \rangle \ \ &=& \ \ 0
\\ \label{e8}
\langle |f(\omega)|^2 \rangle \ \ &=& \ \
\tilde{\rho}(\omega;\mbox{wpk}) \times
\rho(\omega;\delta x)
\end{eqnarray}
where $\tilde{\rho}$ is, up to normalization,
the auto convolution of $\rho(\omega;\mbox{wpk})$,
and therefore equals to the Fourier transform
of $P(t;\mbox{wpk})$, and has roughly the same
width as $\rho(\omega;\mbox{wpk})$.

It is worth noticing that for Lorentzian line shape,
which in general is not necessarily the case,
Eq.(\ref{e8}) implies that $m(t)$ is characterized
by exponential correlations with decay constant $\Gamma/2$.
This leads to decay constant $\Gamma$ for $M(t)$.
The deviation of $\gamma(\delta x)$
form $\Gamma(\delta x)$ in the MBH case
can not be explained by Eq.(\ref{e8}),
since the latter does not distinguish
between the MBH model and the associated
RMBH model.
In order to explain the PID in the MBH case
it is essential to take into account the
non-universal (semiclassical) features of the dynamics.

%%%%%%%%%%%%%%%%%%%%%%%%%%%%%%%%%%%%%%
%%%%%%%%%%%%%%%%%%%%%%%%%%%%%%%%%%%%%%
\section{The semiclassical theory for $P(t)$}

The semiclassical theory of the survival
probability is described within the framework
of wavepacket dynamics in Ref.\cite{heller}.
The short time decay of $c(t;\mbox{wpk})$
reflects the loss of overlap between the
initial and the evolving wavepackets.
On the other hand, due to the (inevitable) proximity
to periodic orbits, the survival amplitude
$c(t;\mbox{wpk})$ have recurrences.
However, because of the (transverse) instability of the
classical motion these recurrences are not complete.
Consequently the long-time decay may be
characterized by the Lyapunov exponent $\gamma_{cl}$.
Possibly, this "Lyapunov decay" is the simplest
example for PID. It is PID because  the size of
the perturbation ($||{\cal H}-{\cal H}_{\tbox{wpk}}||$) 
is not relevant here.

The semiclassical behavior of the survival probability
has reflection in the LDOS structure. A relatively
slow "Lyapunov decay" (due to recurrences) implies
that the LDOS is "scared" \cite{heller}.
Thus the semiclassical LDOS has a `landscape' which
is characterized by the energy scale $\hbar\gamma_{cl}$.
Note that "scarring", in the mesoscopic physics terminology
is called "weak localization" effect.

The above semiclassical picture regarding  $c(t;\mbox{wpk})$
can be extended \cite{wls,prm,vrn}
to the case of $c(t;\delta x)$, provided $\delta x$
is large enough. In the other limit, where $\delta x$
is small, we should be able to use perturbation theory
in order to predict the decay rate. Thus we have
here a {\em clash} of two possibilities:
Having Wigner type decay with $\gamma = \Gamma(\delta x)/\hbar$,
or having non-universal decay (NUD) that reflects the
semiclassical wavepacket dynamics.

The crossover from the perturbative
to the semiclassical regime can be
analyzed \cite{wls,vrn} by looking on
the parametric evolution of $\rho(\omega;\delta x)$.
Depending on $\delta x$ the LDOS $\rho(\omega;\delta x)$
has (in order of increasing perturbation)
either standard perturbative structure,
or core-tail (Lorentzian-like) structure,
or purely non-perturbative structure \cite{rmrk2}.
The width $\Gamma(\delta x)$ of the "core" defines
a `window' through which we can view the
semiclassical `landscape'. This landscape is
typically characterized by $\hbar\gamma_{cl}$ features,
where $\gamma_{cl}$ is related to the classical dynamics.
As $\delta x$  becomes larger, this `window' becomes
wider, and eventually some of the semiclassical
landscape is exposed. Then we say that the LDOS contains
a "non-universal" component \cite{rmrk2}.

%%%%%%%%%%%%%%%%%%%%%%%%%%%%%%%%%%%%%%
%%%%%%%%%%%%%%%%%%%%%%%%%%%%%%%%%%%%%%
\section{The semiclassical theory for $M(t)$}

Whereas Lyapunov decay for $c(t;\delta x)$
is typically a "weak" feature [this is true
for generic systems, whereas billiard systems
constitute an exception], it is not so for $m(t;\delta x)$.
By definition the trajectory of the wavepacket
is reversed, and therefore the short-time decay
due to a loss of wavepacket overlap is avoided.
As a results the perturbation independent
"Lyapunov decay" becomes a predominant feature
(that does not depend on "recurrences").
This Lyapunov PID has been discussed
in \cite{jalabert}.

It is clear however that for small $\delta x$ we can use
perturbation theory in order to predict the decay
rate of $M(t;\delta x)$. The question that naturally arise,
in complete analogy to the $P(t;\delta x)$ case, is
how to determine the border $\delta x_{\tbox{NUD}}$
between the perturbative regime
(where we have Wigner type decay)
and the semiclassical regime (where we have NUD).

The natural identification of $\delta x_{\tbox{NUD}}$
is as the $\delta x$ for which $\Gamma(\delta x)$
becomes equal to $\hbar\gamma_{scl}$.
How $\gamma_{scl}$ is determined?
There are two "mechanisms" that are responsible
to the loss of wavepacket overlap. One is indeed
related to the instability of the classical motion,
while the other is related to the overall energy width
of the wavepacket.

The survival probability $P(t;\delta x)$
can be regarded as a special case of $M(t;\delta x)$,
where the overall energy width of the wavepacket
is the predominant limiting factor in the decay.
the separation between the energy surfaces
of ${\cal H}$ and of ${\cal H}_0$ is proportional
to $\delta x$. Consequently we typically have
$\gamma_{scl}\propto\delta x$.

In the prevailing studies of $M(t;\delta x)$,
one assumes {\em wide} Gaussian wavepackets.
Therefore the separation between the energy
surfaces does not play a major role in the
semiclassical analysis. Rather it is
the instability of the classical motion
that is the predominant limiting factor in the decay.
Therefore one typically expects
to have $\gamma_{scl}\approx\gamma_{cl}$,
which is independent of $\delta x$.

%%%%%%%%%%%%%%%%%%%%%%%%%%%%%%%%%%%%%%
%%%%%%%%%%%%%%%%%%%%%%%%%%%%%%%%%%%%%%
\section{Conclusions}

The above discussed criterion for the identification
of the non-universal regime is in the spirit
of spectral statistics studies \cite{berry}.
In the latter context it is well known
that RMT considerations dominate
the sub $\hbar\gamma_{cl}$ energy scale,
while non-universal corrections dominate
the larger energy scales.

In the present Paper we have identified
the "non-universal" regime for a billiard
related model (MBH).
The border between the perturbative regime
and the non-universal regime in the context
of $P(t;\delta x)$ is $\delta x_{\tbox{NU}}$,
while in the context of $M(t;\delta x)$
it is $\delta x_{\tbox{NUD}}$.

The parametric scales $\delta x_{\tbox{NU}}$
and $\delta x_{\tbox{NUD}}$ are similarly defined,
but there is an important distinction between them.
The first parametric scale marks the exposure
of the semiclassical "landscape": either that
of Eq.(\ref{e3}) or that of Eq.(\ref{e4}).
The second parametric scale, as proved by our
numerical strategy, marks the exposure
of cross correlations between the corresponding
wave amplitudes.

%%%%%%%%%%%%%%%%%%%%%%%%%%%%%%%%%%%%%%%%%%%%%%%%%%%
%%%%%%%%%%%%%%%%%%%%%%%%%%%%%%%%%%%%%%%%%%%%%%%%%%%
\appendix
\section{The $\delta x_{\tbox{NU}}$ for billiards}

The non-universal regime for billiard systems
has been identified in \cite{wls,prm}.
Here we would like to complete missing steps
in the generalization of this result.
We use the same notations as in \cite{wls,prm}.

In the general case \cite{prm} the bandprofile
of the $\mbf{B}$ matrix is determined by the 
semiclassical formula \cite{mario}
\begin{eqnarray} 
\left\langle\left|\mbf{B}_{nm}
\right|^2\right\rangle
\ \ \approx \ \
\frac{\Delta}{2\pi\hbar} \
\tilde{C}\left(\frac{E_n{-}E_m}{\hbar}\right)
\end{eqnarray}
where $\Delta \propto 1/k^{d{-}2}$
is the mean level spacing, $k$ is the wavenumber,
and $d=2$ is the dimensionality of the Billiard system.
The power spectrum of the motion
\begin{eqnarray}
\tilde{C}(\omega) \ = \ \mbox{const} \times k^{3+g}/\omega^{g}
\end{eqnarray}
is the Fourier transform of a classical correlation function.
Here $g=0$ corresponds to strong chaos assumptions,
while $0<g<1$ is more appropriate for our type of
system due to the bouncing ball effect.
The width of the LDOS is determined using 
a procedure which is explained in \cite{wls}, 
leading to Eq.(9) there. Namely,  
\begin{eqnarray} \label{eq_G}
\Gamma(\delta x) \ = \ \Delta \times (\delta x/\delta x_c)^{2/(1+g)}
\end{eqnarray}
where $\delta x_c \propto k^{-((1{-}g)+(1{+}g)d)/2}$
is the generalization of Eq.(8) of \cite{wls}.

Form Eq.(\ref{eq_G}) it is clear that $\delta x_c$
should be interpreted as the deformation
which is needed in order to mix neighboring levels.
In the standard perturbative regime
($\delta x \ll \delta x_c$) first order
perturbation theory is valid as a global approximation.
Otherwise, if $\delta x > \delta x_c$, we should distinguish
between a non-perturbative "core" of width $\Gamma$,
and perturbative "tails" that lay outside of it.

The expression for $\Gamma$ can be re-written as
\begin{eqnarray}
\Gamma(\delta x) \ \approx \ \hbar\gamma_{cl} \times (k\delta x)^{2/(1{+}g)}
\end{eqnarray}
where $\gamma_{cl} \propto k$ is roughly the
inverse of the ballistic time.
In our numerical analysis \cite{wisniacki}
we find that $\Gamma\approx 0.36 k^2 \times \delta x$,
corresponding to $g=1$.
The non universal scale $\delta x_{\tbox{NU}}$,
as well as $\delta x_{\tbox{NUD}}$,
are determined by the requirement
$\Gamma(\delta x) = \hbar\gamma_{cl}$.
Hence we get Eq.(\ref{eq_DeB}), which
holds {\em irrespective} of the $g$ value.
The latter claim has been stated in \cite{wls}
without a proof.

%%%%%%%%%%%%%%%%%%%%%%%%%%%%%%%%%%%%%%%%%%%%%%%%%%%%%%%%%%%%%%
%%%%%%%%%%%%%%%%%%%%%%%%%%%%%%%%%%%%%%%%%%%%%%%%%%%%%%%%%%%%%%

\ \\

\noindent
{\bf Acknowledgments:} We thank Tsampikos Kottos
and Eduardo Vergini for useful discussions.
DAW gratefully acknowledges support from CONICET
(Argentina). Research grants by  CONICET and ECOS-SeTCIP.

%%%%%%%%%%%%%%%%%%%%%%%%%%%%%%%%%%%%%%%%%%%%%%%%%%%%%%%%%%%%%%%%%
%%%%%%%%%%%%%%%%%%%%%%%%%%%%%%%%%%%%%%%%%%%%%%%%%%%%%%%%%%%%%%%%%

%%%%%%%%%%%%%%%%%%%%%%%%%%%%%%%%%%%%%%%%%%%%%%%%%%%%%%%%%%%%%%%%
%%%%%%%%%%%%%%%%%%%%%%%%%%%%%%%%%%%%%%%%%%%%%%%%%%%%%%%%%%%%%%%%

\ \\
\ \\

%%%%%%%%%%%%%%%%%%%%%%%%%%%%%%%%%%%%%%%%%%%%%%%%%%%%%%%%%%
%%%%%%%%%%%%%%%%%%%%%%%%%%%%%%%%%%%%%%%%%%%%%%%%%%%%%%%%%%
%\begin{figure}[h]
\centerline{\epsfig{figure=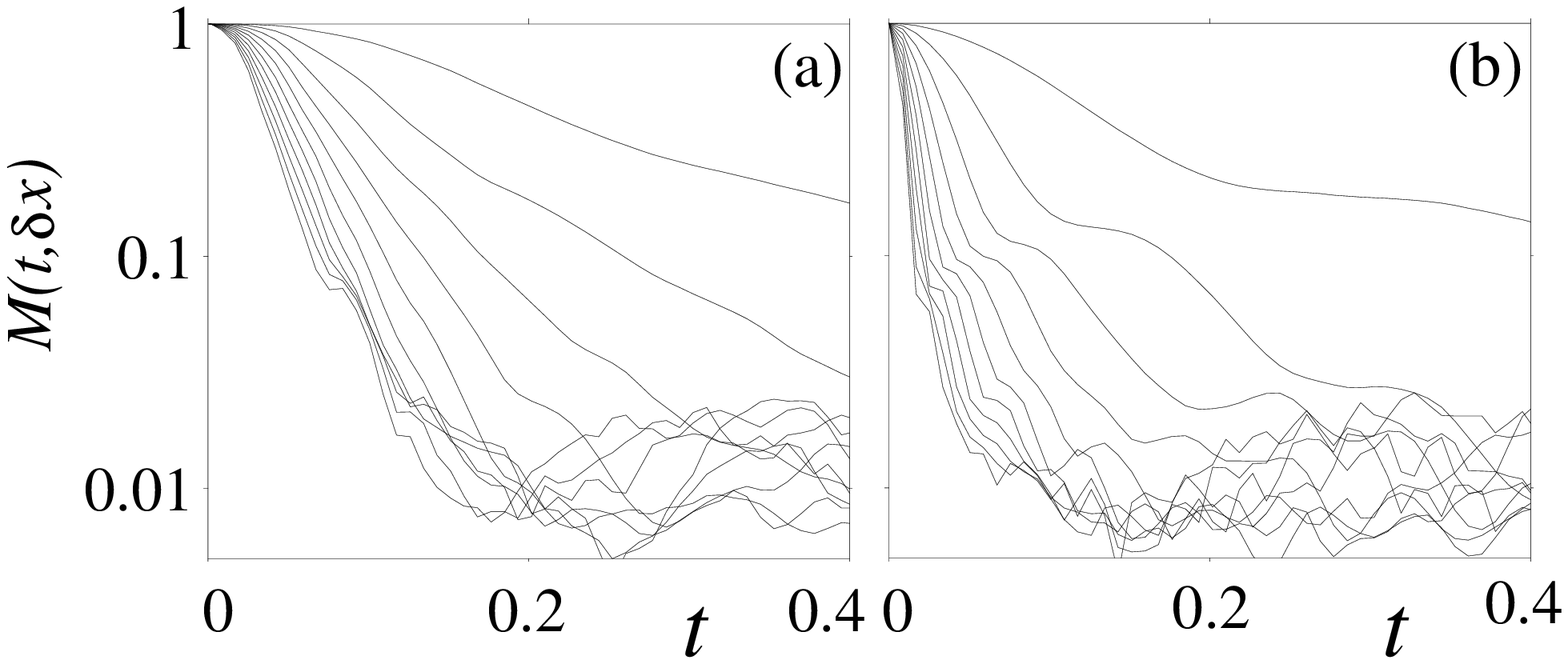,width=\hsize}}
%\vspace{.1in}
{\footnotesize FIG1:
(a) The decay of $M(t;\delta x)$ in the MBH case.
(b) The same for the randomized MBH (RMBH).
We use dimensionless units of time that correspond
to stadium billiard with straight edge $x_0=1$,
particle with mass $m=1/2$, wavenumber $k\sim50$,
and $\hbar=1$.
The values of the perturbation strength are
(from the top curve to bottom):
$\delta x=0.0125*i$ with $i=1,\cdots,11$.}
%\end{figure}
%%%%%%%%%%%%%%%%%%%%%%%%%%%%%%%%%%%%%%%%%%%%%%%%%%%%%%%%%%
%%%%%%%%%%%%%%%%%%%%%%%%%%%%%%%%%%%%%%%%%%%%%%%%%%%%%%%%%%

\ \\

%%%%%%%%%%%%%%%%%%%%%%%%%%%%%%%%%%%%%%%%%%%%%%%%%%%%%%%%%%
%%%%%%%%%%%%%%%%%%%%%%%%%%%%%%%%%%%%%%%%%%%%%%%%%%%%%%%%%%
%\begin{figure}[h]
\centerline{\epsfig{figure=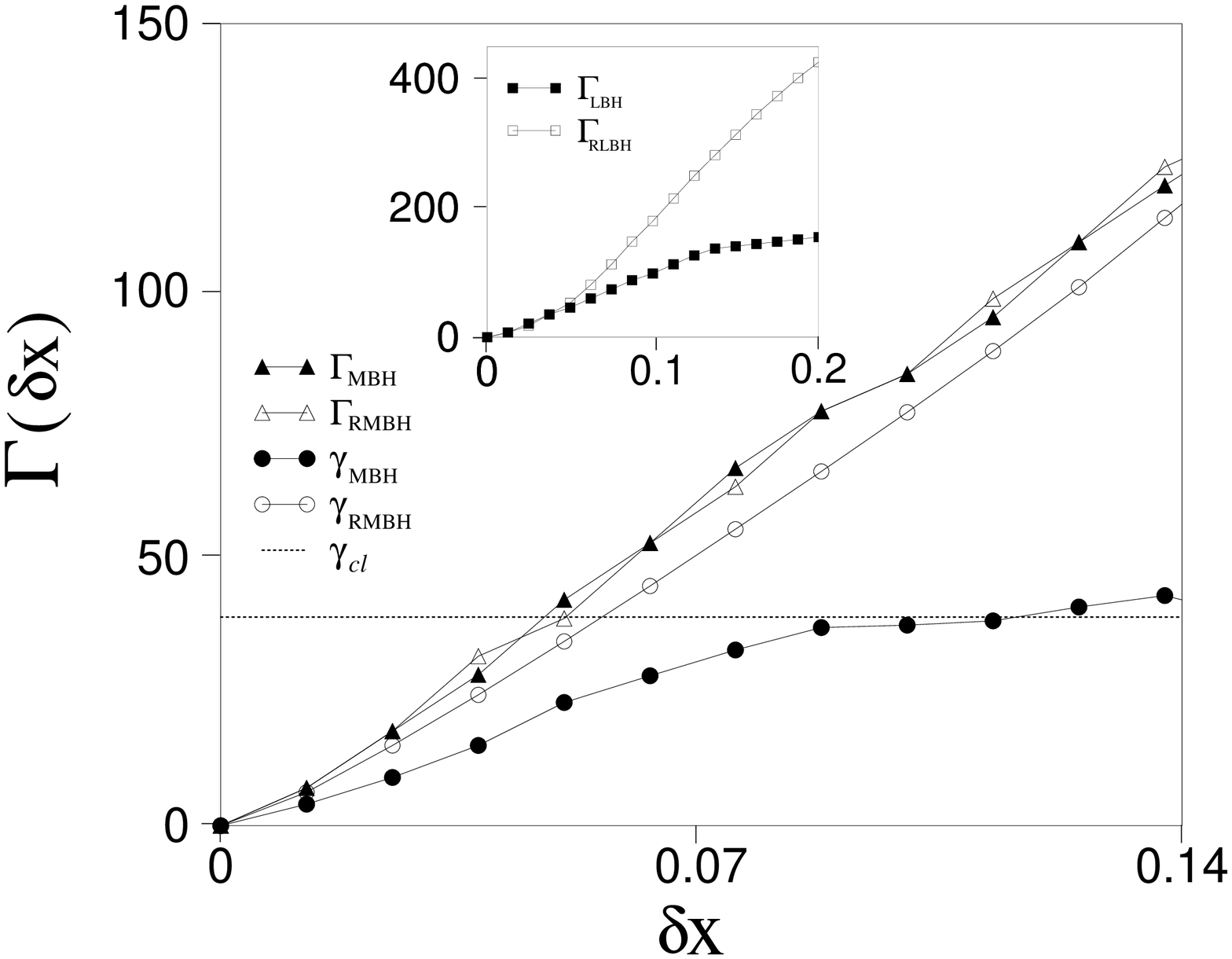,width=\hsize}}
%\vspace{.1in}
{\footnotesize FIG2:
The LDOS width $\Gamma$, and the decay constant $\gamma$
from the MBH/RMBH simulations.
The dotted line is the classical Lyapunov exponent.
The inset is $\Gamma$ in the LBH/RLBH case.}
%\end{figure}
%%%%%%%%%%%%%%%%%%%%%%%%%%%%%%%%%%%%%%%%%%%%%%%%%%%%%%%%%%
%%%%%%%%%%%%%%%%%%%%%%%%%%%%%%%%%%%%%%%%%%%%%%%%%%%%%%%%%%


\begin{thebibliography}{99}
\bibitem{rmrk0}
We use $\hbar=1$ units, but later is some
formulas we include $\hbar$ in order to make
it easy for the reader to make comparison
with other publications.
\bibitem{rmrkP}
The term "parametric evolution" refers to the
$\delta x$ dependence of LDOS profile,
which starts as a delta distribution
at $\delta x=0$. It can be formally obtained
as the solution of a Schrodinger-like equation,
where $\delta x$ plays the role of fictitious time.
\bibitem{rmrk1}
The saturation of $\Gamma(\delta x)$ is
just "postponed" by the cutoff procedure.
This means that by looking on large
enough $\delta x$ range we can still
observe reminisces of saturation.
For stronger cutoff, a larger range
is required to see deviation from the
approximately linear behavior
of~$\Gamma(\delta x)$.
\bibitem{rmrk2}
If the perturbative tails effectively
disappear, then we say that the LDOS has
a purely "non-perturbative" structure.
For large enough $\delta x$ the non-perturbative
structure becomes purely semiclassical.
\end{thebibliography}
\end{document}